# Microfluidics for Biofabrication

*Shoji Takeuchi - University of Tokyo*

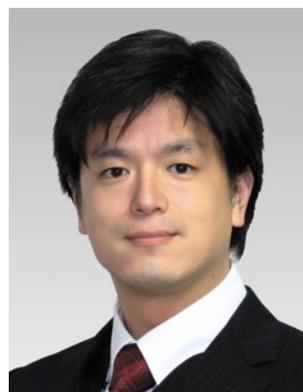

## Biography

*Shoji Takeuchi received his B.E., M.E., and Dr. Eng. degrees in mechanical engineering from the University of Tokyo, Tokyo, Japan in 1995, 1997, and 2000, respectively. He is currently a professor in and the Director of the Center for International Research on Integrative Biomedical Systems (CIBiS), Institute of Industrial Science (IIS) at the University of Tokyo. He has authored more than 150 peer-reviewed publications and filed over 70 patents. He has been recognized with numerous honors, including MEXT Young Scientists' Prize in 2008, the JSPS prize in 2010, and the ACS Analytical Chemistry Young Innovator Awards in 2015. His current research interests include 3D tissue fabrication, implantable devices, artificial cells/lipid bilayer systems, and biohybrid MEMS.*

## Summary

Living materials show functions superior to artificial ones in terms of the ability to recognize/produce biomolecules. For example, cells, particularly membrane proteins, achieve a highly sensitive and selective detection of several chemicals at the molecular level. They can even distinguish between different chemical moieties found in various odorants. When a certain ion channel is activated by a single ligand molecule, it generates a small current of few pico-amperes under a 100 mV membrane potential. This reaction means that the single molecular signal is amplified with approximately 107 ions per second in living systems. Therefore, the system can be regarded as a transistor with an excellent amplifier. The motivation here is to use this function in an engineered system by reconstructing the biological structure using microfluidic technology.

One of my talks pertains to the reconstruction of a cell membrane integrated with membrane proteins in a microfluidic chip. As mentioned above, membrane proteins play very important roles in cells. They are also useful in various industrial fields, including next-generation diagnosis techniques, drug discovery, and so on. We have applied membrane proteins for highly sensitive and selective chemical sensors in a previous study (PNAS 107: 15340-15344, 2010), wherein we used a living cell expressing odorant receptors as the sensor that could distinguish multiple-target chemicals. This cell-based sensor can be integrated with a humanoid robot without any noise reduction systems. The sensor can be compact, and thus, easy to be incorporated into a portable device useful for environmental monitoring.

Moreover, we developed a simple method to use or analyze membrane proteins by incorporating them into artificially formed planar lipid bilayers. Lipid bilayers produced by conventional methods are often fragile, unsteady, and difficult to be reproduced. These weak properties reduce their usefulness in high-throughput systems. We thus developed a reproducible method, called the "droplet contacting method" for forming planar bilayers using simple fluidic control (Anal. Chem. 78:8169-8174, 2006) (Fig. 1). This method is very simple and reproducible. Hence, it is now applied in various types of devices for membrane protein analysis (Anal. Chem. 81:9866-9870, 2009, Small 6:2100-2104, 2010, LOC 11:2485.2487, 2011, etc.) and nowadays recognized as the "droplet interface bilayer (DIB)" widely used by many research groups. We are applying this lipid bilayer system to various fields, including chemical sensing (JACS 133:8474-8477, 2011), protein synthesis (Anal. Chem. 83:3186-3191, 2011), drug kinetics studies (Scientific Reports in press, LOC 12:702-704, 2012), and artificial cells studies (Nature Chemistry 8:881–889, 2016, Angew. Chem. Int. Ed., 48: 6533-6537, 2009, JACS 129:12608-12609, 2007, APL 96: 083701, 2010).

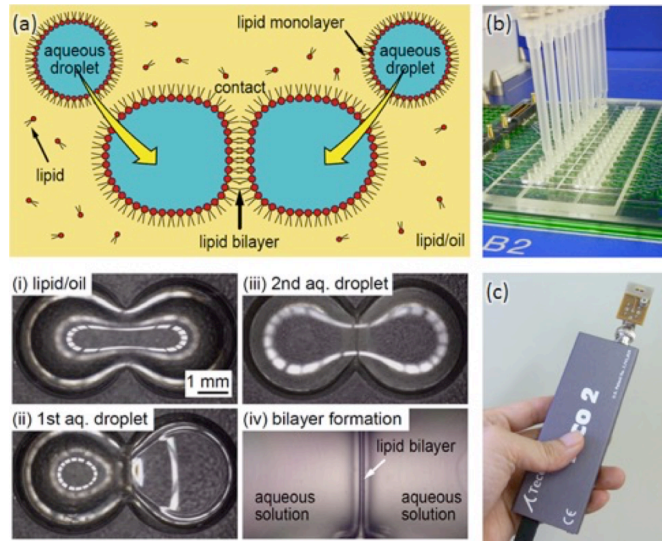

*Fig. 1: (a) Droplet contacting method: when two water droplets are in oil containing amphiphilic molecules (phospholipids), the monolayer spontaneously assembles at the interface between the water and the oil. Once the two interfaces come into contact with each other, they form a lipid bilayer. (b) An array of the lipid bilayer formed by DCM for a high-throughput ion channel recording. (c) Portable system for recording the activity of membrane proteins reconstituted into the lipid bilayer membrane.*

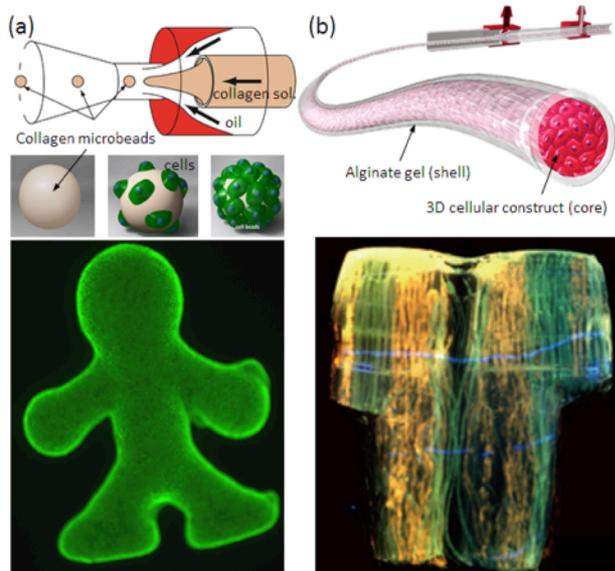

*Fig. 2: (a) Axisymmetric flow focusing device (AFFD) for producing monodisperse collagen hydrogel beads. These beads are used as substrates for the 3D cell culture. These cell-laden hydrogel beads (cell beads) are molded into a highly dense 3D macroscopic tissue. (b) Double coaxial microfluidic device for producing a core-shell hydrogel microfiber. The cells in the core can grow and migrate with an ECM-rich microenvironment and eventually form a fiber-based 3D tissue. The fiber is easy to handle and useful for a higher-order assembly, such as a woven fabric.*

The other talk describes microfluidic-based approaches for the rapid construction of a large-scale three-dimensional (3D) tissue that mimics a microscopic tissue structure in vivo. We demonstrated a bottom–up tissue construction method using different types of cellular modules that serve as building blocks for thick and dense 3D tissues (e.g. cell beads and cell fibers).

We prepare the cell-laden hydrogel beads (cell beads) by using an axisymmetric flow focusing device (AFFD) to form monodisperse collagen beads that can be used as the substrate of the cells. The cells can adhere, grow, and migrate on the collagen gel beads. These cell beads are then molded into a designed polydimethylsiloxane chamber to form macroscopic 3D tissue structures. In the mold, the cell beads can adhere to each other via the cells coated on the collagen gel beads. The cells also migrate and grow into collagen gel beads with contraction and decomposition of the collagen to ultimately form the macroscopic tissues (Fig. 2a). Finally, the fabricated tissue can be obtained by releasing from the mold. The bead-stacking structure allows nutrients to reach the cells located in the center of the tissue, thereby preventing necrosis during tissue formation for more than a day. We found that this property was not observed with cell aggregates. This approach enables the rapid and reproducible construction of large-scale 3D tissues with a complex microstructure.

As regards the fiber-shaped cellular building units, we established a method to produce core-shell hydrogel microfibers by using a double-coaxial microfluidic device (Fig. 2b). The core surrounded by the shell of the mechanically stable Ca-alginate hydrogel can be supramolecules (Angew. Chime 51: 1553-1557, 2012, and 51: 7942-7947, 2012), bacterial cellulose fibers (Biomaterials, 34: 2421-2427, 2012), and cell-containing ECM proteins (Nature Materials, 12: 584-590, 2013). The cells cultured in the core of the fiber show excellent intrinsic functions. In combination with myocytes, endothelial, and nerve cells, the cells showed the contractile motion of the myocyte cell fiber, tube formation of the endothelial cell fibers, and synaptic connections of the nerve cell fiber, respectively. A higher-order assembly of fiber-shaped 3D cellular constructs can be performed by using microfluidic handling. Mechanical weaving of cell fibers with our lab-made microfluidic weaving machine particularly provides a woven "cell fabric" composed of three different cell fibers.

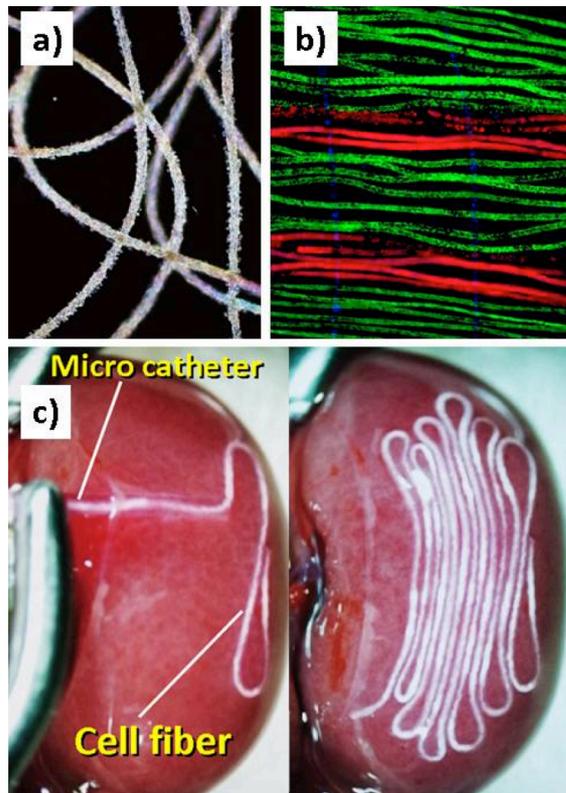

Fig. 3: (a) A meter-long, 200 μm-diameter cell fiber (b) A centimeter-scale woven macroscopic tissue using three different cell fibers assembled by a weaving machine. (c) Images of the implantation of a 20 cm-long beta cell-laden fiber into the subrenal capsular space of a recipient diabetic mouse during (left) and after (right) the implantation. (Nature Materials 2013)

As a practical application, fiber-encapsulating beta-cells were used for the implantation of diabetic mice. We first prepared a pancreatic islet cell fiber using primary rat dissociated islet cells. The primary islet cell fiber was then injected into the subrenal capsular space of a diabetic mouse using a microcatheter. As a result, we found that the implanted fibers normalized blood glucose concentrations. Moreover, the fiber removal caused an immediate reappearance of hyperglycemia in all mice. These results indicate that our cell fiber is useful for applications in medical transplantation using regenerated glucose-responsive insulin-secreting cells, and could potentially be used as a removable graft when the cell fiber is covered with a stable shell in vivo.